\documentclass[conference,a4paper]{APSIPA2021}
\usepackage{multirow}
\usepackage{graphicx}
\usepackage{amsmath}
\usepackage[psamsfonts]{amssymb}
\usepackage{amsxtra}
\usepackage{threeparttable}

\usepackage{makecell}
\newcolumntype{?}{!{\vrule width 1.5pt}}

\newcommand\blfootnote[1]{%
  \begingroup
  \renewcommand\thefootnote{}\footnote{#1}%
  \addtocounter{footnote}{-1}%
  \endgroup
}

\begin{document}

\title{Amino Acid Classification in 2D NMR Spectra via Acoustic Signal Embeddings}

\author{%
\authorblockN{%
Jia Qi Yip\authorrefmark{4}\authorrefmark{2}, Dianwen Ng\authorrefmark{4}\authorrefmark{2}, Bin Ma\authorrefmark{4}, Konstantin Pervushin\authorrefmark{2} and Eng Siong Chng\authorrefmark{2}
}
\authorblockA{%
\authorrefmark{4}
Alibaba Group
}
\authorblockA{%
\authorrefmark{2}
Nanyang Technological University, Singapore\\
E-mail: jiaqi006@e.ntu.edu.sg}
}

\maketitle
\thispagestyle{empty}

\blfootnote{This work was supported by Alibaba Group through Alibaba Innovative Research (AIR) Program and Alibaba-NTU Singapore Joint Research Institute (JRI), Nanyang Technological University, Singapore.}

\begin{abstract}
Nuclear Magnetic Resonance (NMR) is used in structural biology to experimentally determine the structure of proteins, which is used in many areas of biology and is an important part of drug development. Unfortunately, NMR data can cost thousands of dollars per sample to collect and it can take a specialist weeks to assign the observed resonances to specific chemical groups. There has thus been growing interest in the NMR community to use deep learning to automate NMR data annotation. Due to similarities between NMR and audio data, we propose that methods used in acoustic signal processing can be applied to NMR as well. Using a simulated amino acid dataset, we show that by swapping out filter banks with a trainable convolutional encoder, acoustic signal embeddings from speaker verification models can be used for amino acid classification in 2D NMR spectra by treating each amino acid as a unique speaker. On an NMR dataset comparable in size with of 46 hours of audio, we achieve a classification performance of 97.7\% on a 20-class problem. We also achieve a 23\% relative improvement by using an acoustic embedding model compared to an existing NMR-based model.
\end{abstract}

\section{Introduction}
Nuclear Magnetic Resonance (NMR) is an instrumental technique used to determine the structure of molecules based on the bulk physical interaction between the strong magnetic fields of atoms and an external weak oscillating magnetic field. It is unique among other instrumental techniques in its ability to determine the 3D structure of proteins in solution --- the natural environment of proteins in the human body~\cite{wuthrichlecture}. NMR has a wide variety of practical applications especially in chemistry and structural biology. Determining the structure of proteins in solution is a critical task as the function of a protein is primarily determined by its structure. It is thus necessary for many applications within the life sciences, such as drug development. 

Much of the recent research in deep learning for NMR is motivated by a need to reduce the amount of labor needed in the analysis of NMR experiments. A single NMR experiment can take many days of data acquisition followed by a mostly manual data analysis. A number of methods have already been developed to tackle this challenge from various angles. To reduce the instrument time needed for data collection, spectral reconstruction from non-uniform sampling~\cite{fidnet}\cite{Hansenfid}\cite{AcclNMR} and spectral de-noising~\cite{nmrdenoise} methods have been developed. For faster and easier data analysis, work has been done on chemical shift prediction~\cite{nmrdlreview}, automated peak picking~\cite{deeppick}, spectral deconvolution for Diffusion-Ordered NMR Spectroscopy~\cite{DOSY} and even direct structure prediction from NMR chemical shifts~\cite{CSI-LSTM}\cite{spec2struct}. These methods either utilize simple models based on common neural network motifs such as CNN~\cite{AcclNMR}, LSTMs~\cite{Hansenfid}\cite{CSI-LSTM} and even transformers~\cite{transformerNMR}. A few time-domain methods have been proposed for spectral reconstruction~\cite{fidnet} and peak picking~\cite{deeppick}. 

However, the availability of training data remains an impediment in developing deep learning methods for NMR. The training of the existing NMR models rely on unrealistic synthetic datasets where data is generated by the simple superposition of randomised exponential functions~\cite{AcclNMR}\cite{fidnet}. This is because it is prohibitively expensive to acquire NMR data at a scale that is sufficient for the training of these models, limiting the types of models that can be trained. Furthermore, one of the defining features of NMR spectra is the coupling pattern between various nuclei in the molecule, which requires taking into account the global context of the molecule. This cannot be easily achieved through the current bottom up approach. 

Additionally, while the aforementioned deep learning approaches to tackling various challenges in NMR have been successful, the pace of research is slower since the NMR community is smaller and has fewer people working on the problem. Given that both NMR signals and acoustic signals are time-domain, it might be more efficient for NMR researchers to utilize models and strategies that have already been developed in acoustic signal processing.

In this paper we show that it is possible to use acoustic signal processing models directly on NMR data to achieve improved performance on a classification task without the need to design new models. This is inspired by the fact that NMR models that have achieved good performance so far have architectures that share structural similarities with audio processing models. For example, \cite{deeppick} has a deep 1D convolutional architecture, while~\cite{fidnet} was explicitly based on WaveNet~\cite{wavenet}. To acquire enough realistic training data to train these models we propose a novel procedure making use of physics-based NMR simulations which can produce essentially all types of NMR experiments and can be applied to any compound. Using our proposed procedure we simulate a time-domain amino acid classification dataset comparable in size to 46 hours of dual-channel 8k audio.

\section{Our Approach}
\label{approach}
In this section, we discuss the similarities between audio and NMR data to establish the plausibility of using acoustic signal processing models on NMR data and the model architectures used in our experiments. Thus, each amino acid can be thought of as a ``speaker'' and each NMR experiment as a different utterance, turning our NMR classification problem into a speaker identification problem. We then discuss how we use the speaker embeddings of pre-existing speaker verification methods available on the Speechbrain framework~\cite{speechbrain}, x-vector~\cite{xvec} and ECAPA-TDNN~\cite{ecapa}, as well as an existing NMR-based classification model, DEEP Picker~\cite{deeppick}, to perform the classification. Details of our amino acid classification dataset and simulation methodology will be discussed in section~\ref{dataset}.

\subsection{The similarities between NMR data and Audio}
In this work we focus on a particular type of NMR experiment, the 2 dimensional Total Correlation Spectroscopy (TOCSY), which detects the frequency correlations between each proton and every other proton in the molecule. It can be thought of as a 2D interferogram, with $n$ transients each containing $m$ data points.\setlength{\parskip}{2pt}

Although the NMR interferogram consists of a collection of independent transients and is typically processed in a higher dimensional manner, the data can also be processed as a single dimension by simply stacking the transients one after another in order to form a single long sequence. This is merely a rearrangement of data points and should not change the content of the signal. This transformation can be described as the transpose of an arbitrary signal $S = \begin{bmatrix} x_{1} \\ \vdots \\ x_{n} \end{bmatrix}$ where $x_{n}\in \mathbb{R}^{m}$ is a row vector.\setlength{\parskip}{4pt}

\begin{figure}[!htbp]
\begin{center}
\includegraphics[width=85mm]{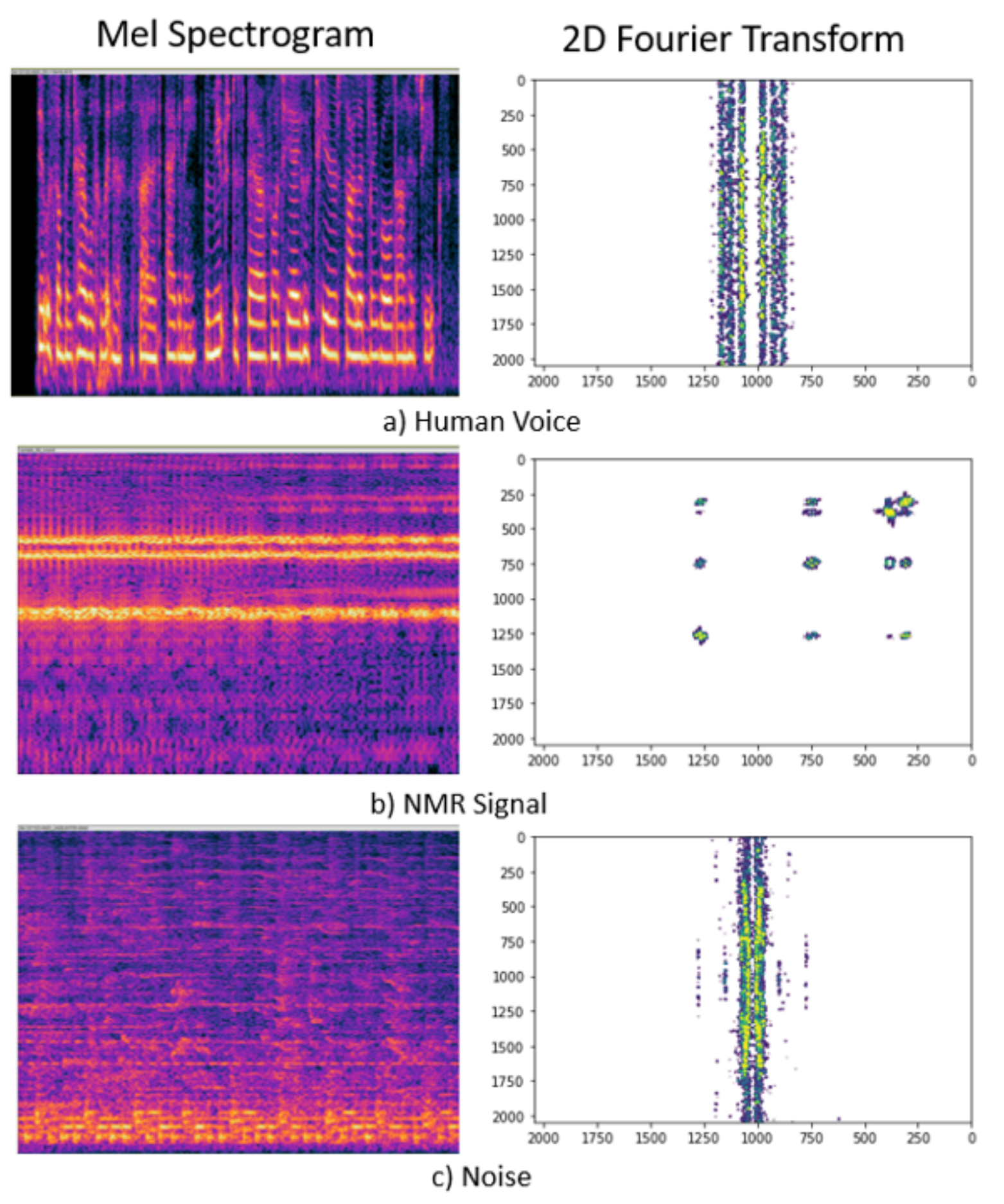}
\end{center}
\caption{Comparison of Audio Data and NMR data using two different visualization techniques. The human voice sample and noise sample are obtained from the 8k dev set of the Libri2Mix Corpus~\cite{librimix}. Mel spectrograms plotted using Audacity~\cite{audacity}, while the 2D Fourier transform is plotted according to standard procedure in our own python implementation.}
\label{nmr-speech-comparison}
\end{figure}

By taking the transpose of the NMR signal we are then also able to plot the Mel spectrogram of an NMR signal as shown in Fig.~\ref{nmr-speech-comparison}, which helps us visualize how NMR signals can be treated as audio signals. We also plot in Fig.~\ref{nmr-speech-comparison} the result of 2D Fourier transform on the transpose of a regular speech signal by taking a window $x_{n}=256$ to align with the NMR signals and truncating the last few data points. The 2D Fourier transform is the standard data processing technique used for visualizing the NMR interferogram.

In Fig.~\ref{nmr-speech-comparison} we observe that the Mel spectrogram representation of the NMR signal has high energy bands where it resembles the speech signal, albeit without the harmonic structures found naturally in human voices. We hypothesize that the presence of these high energy bands provide enough similarity to human speech to allow the use of acoustic signal embeddings for classifying these NMR.\setlength{\parskip}{2pt}

Looking at the 2D Fourier transforms, we observe that there are significant differences between NMR spectra and the audio signals. This is to be expected as the audio used in this case is single channel and real valued, so the 2D Fourier transform operation is missing complex information from the input signals. However, we still note that the noise signal has an observably different pattern from the human voice signal, just as they are different in the Mel spectrogram representation.

\subsection{Model Architecture}

\begin{figure}[!htbp]
\begin{center}
\vspace*{-3pt}
\includegraphics[width=87mm]{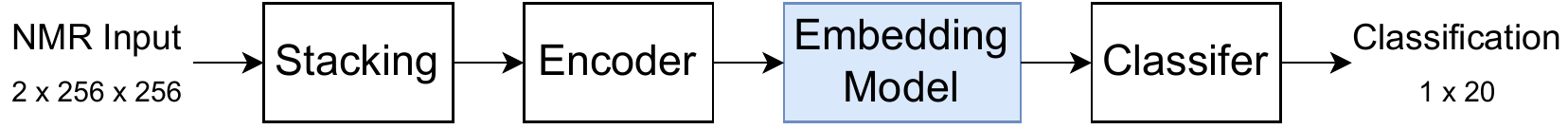}
\end{center}
\caption{The overall architecture of the model used in this study. We explore the impact of using different embedding models on the classification performance of the overall model. The 2 channel NMR input consists of 2 complex channels, which are also converted into 4 real channels by the dataloader before the stacking process.}
\label{model-architecture}
\end{figure}

We adopt a standard classifier pipeline which will allow us to swap out different embedding models within the same structure to test their relative performance\footnote{Code available at github.com/Yip-Jia-Qi/NMR-Classification}. The overall architecture of the model is shown in Fig.~\ref{model-architecture}. The stacking operation is the transpose of the NMR data as discussed in \ref{approach}, while the Encoder makes use of either a standard filter bank (which produces 80-dimensional Mel Filters from a 25 ms window with a 10 ms frame shift) or a single 1D convolutional layer with a kernel size of 256, inspired by~\cite{AMS-SE}. The classifier is a single linear layer with a softmax activation function. The overall model can be described as follows:

\begin{equation}
    S_{emb} = Embedding(Encoder(S_{NMR}))
\end{equation}
\begin{equation}
    P_{1\times20} = Softmax(\sum_{i=1}^{k}{Linear(S_{emb_{i}})})
\end{equation}
where $k = 4$ in this case as our data has 4 channels from converting 2 complex channels in the NMR data into 4 real channels. 

Using the proposed architecture, we experiment with three different embedding models, two which are acoustic signal embedding models and one which is an NMR-based classification model:
\begin{itemize}
    \item{\textbf{X-Vector}. The x-vector model~\cite{xvec} is a common baseline used in speaker verification work. It makes use of a deep convolution architecture with a statistical pooling layer. We train the model embeddings as implemented on Speechbrain with no modifications.}\setlength{\parskip}{5pt}
    
    \item{\textbf{ECAPA-TDNN-S/L}. The Emphasized Channel Attention, Propagation, and Aggregation - Time Delay Neural Network (ECAPA-TDNN) model~\cite{ecapa} is a recently proposed model readily available on Speechbrain. It makes use of multiple strategies such as a deep convolutional architecture with skip connections, grouped convolutions, squeeze excitation blocks, and attentive statistics pooling and has been shown to outperform x-vector models in speaker verification. We train two variations of the model, S and L, Which are configured as reported in~\cite{ecapa} with channel size in the intermediate layers of the model set to 512 and 1024 respectively.}\setlength{\parskip}{5pt}
    
    \item{\textbf{DEEP Picker}. The Deep nEural nEtwork Peak Picker (DEEP Picker)~\cite{deeppick} model is a non-speech model specially designed to work on NMR tasks. As the model was originally implemented in C++, we rewrote the model in PyTorch. The model was originally designed for NMR peak-picking, which can be easily adapted for this classification task. The model consists of 7 layers of 1D convolutions. Kernel and filter sizes\footnote{The dimensions of the model are (11, 1, 11, 1, 1, 11, 1) for the kernels and (40, 20, 10, 20, 10, 30, 18) for the filters} are implemented as described in~\cite{deeppick}. The only modification we make is to the design of the final layer, where we have the model output 20 classes, instead of 3 classes in the original model.}
\end{itemize}

\section{Dataset}
\label{dataset}

\begin{figure}[bhtp]
\begin{center}
\includegraphics[width=87mm]{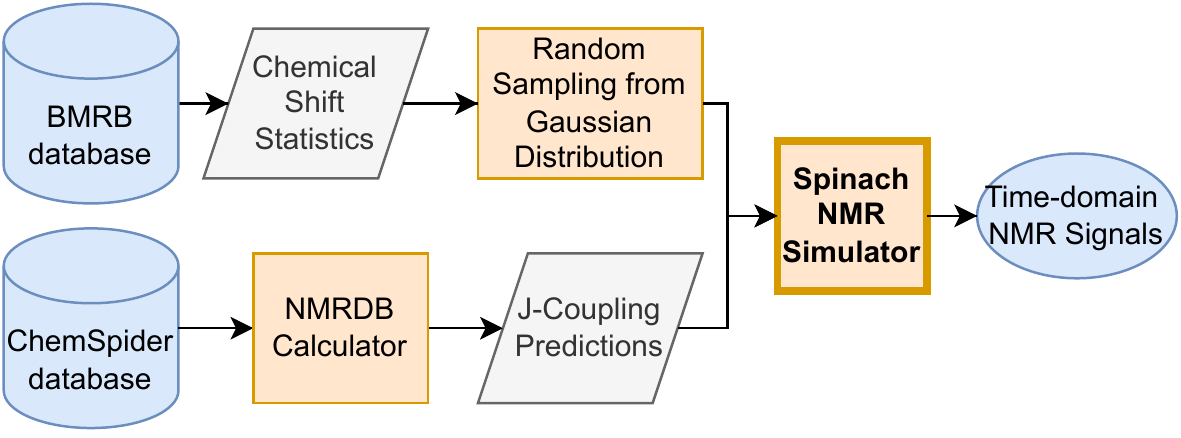}
\end{center}
\caption{Flowchart outlining the simulation procedure. BMRB stands for the Biological Magnetic Resonance Data Bank. Chemical shifts are the resonance frequencies of an amino acid. There is some natural variation in these chemical shifts which are captured in the statistics on BMRB. ChemSpider is a chemical database where we can obtain chemical structures while NMRDB is a tool for obtaining J-coupling values from chemical structures. J-coupling values are characteristics of an amino acid and depend on the chemical bonds between each atom in a molecule. Both are necessary inputs for the Spinach NMR Simulator.}
\label{DataGenerationDiagram}
\end{figure}

\subsection{Dataset Generation Procedure}
Our procedure for the generation of realistic NMR spectra requires a number of tools\footnote{Code available at github.com/Yip-Jia-Qi/Spinach-Amino-Acid-Simulation}. The overall flow of data and calculations is depicted in Fig.~\ref{DataGenerationDiagram}. At the core of it, is the Spinach library~\cite{HOGBEN}, which is able to accurately and efficiently model the physical evolution of the spin system of an amino acid molecule to generate realistic solution NMR spectra. The Spinach library's physics model requires a detailed empirical specification of the physical attributes of the target molecule, which is assembled from a few sources. Firstly, chemical shift statistics are collected from the Biological Magnetic Resonance Data Bank (BMRB)~\cite{BMRB}. The chemical shifts of the samples generated for our dataset are then drawn from a Gaussian distribution reported in the BMRB chemical shift statistics. Next, molecular structures are downloaded from ChemSpider~\cite{chemspider} and plugged into NMRDB~\cite{nmrdb1}\cite{nmrdb2}\cite{nmrdb3} to obtain the typical values for J-couplings to be paired with the chemical shift statistics. Briefly, the J-coupling values depend on the the chemical bonds between each atom in a molecule. In our simulations the J-coupling values are kept constant and only the chemical shifts change from sample to sample. Finally, we input the symmetry of each amino acid (i.e. duplicate atoms, such as the three hydrogen in $CH_{3}$) into Spinach by hand.

This dataset simulation approach has two benefits. Firstly, once the spin system has been specified, we will be able to simulate any desired NMR experiment including all common multidimensional experiments used in structural biology. Secondly, the number of samples and resolution of the simulated spectra can also be arbitrarily specified, avoiding issues arising from mismatched resolutions. With this, a large dataset can be generated quickly for almost any NMR task without resorting to the simplifications used in previous work~\cite{AcclNMR}\cite{fidnet}.

\textbf{Spinach}. NMR data consists of the resonances of nuclear spins within a molecule. In a real-world experiment, the instrument detects the oscillations in magnetic field of individual spins at their respective resonant frequencies. The instrument seeks to track the subject molecules' response to external radio-frequency controls as it evolves with time. Since the NMR experiment exists in a well-controlled environment and the physics is well understood at the scale we are concerned with, the evolution of the spin system can in principle be computationally modelled in a deterministic manner --- as long as the relevant physical parameters of the subject molecule (i.e. the spin system) can be properly specified.\setlength{\parskip}{2pt}

Time-domain NMR simulation packages such as Spinach solve the Liouville-von Neumann equation for the spin-density operator in order to obtain the observable magnetization as the simulation moves forward in time~\cite{SpinachTut}. The oscillation of the observable magnetization is what gives rise to the time-domain NMR signal as described in (\ref{Liouvillian}) and (\ref{mag}).

\begin{equation}\label{Liouvillian}
    \frac{\partial }{\partial t}\hat{\rho}(t) = -i \widehat{\hat{L}}(t)\hat{\rho}(t)
\end{equation}
\setlength{\parskip}{5pt}
\begin{equation}\label{mag}
    m(t) = \left \langle \widehat{m}|\hat{p}(t) \right \rangle
\end{equation}
where $\hat{\rho}(t)$ is the spin-density operator represented as a matrix containing information about the spin-state of the molecule, $\widehat{\hat{L}}$ is a matrix representing the relationships between the atoms in the molecule (also known as the Liouvillian) and $\widehat{m}$ is a projector of the observable magnetization~\cite{SpinachTut}. We may solve (\ref{Liouvillian}) by calculating the exponential of $\widehat{\hat{L}}$ as per (\ref{LVSoln}):

\begin{equation}\label{LVSoln}
    \hat{\rho}(t + dt) = \textup{exp}\left [-i \widehat{\hat{L}}(t)dt\right ]\hat{\rho}(t)
\end{equation}

In practice, these equations involve matrices that are quite large, resulting in an astronomical order of computation. Thus, some assumptions have to be made to make the problem tractable. These assumptions can be made with minimal trade-off in accuracy due to well-designed algebraic shortcuts implemented in Spinach~\cite{HOGBEN}. When given the appropriate physical parameters of the system, Spinach is able to perform accurate, physics-based computational modelling of the spin system evolution in order to generate a number of different experiments~\cite{SpinachTut}. The outputs of this model are widely accepted to be SOTA and work has already been done to establish that the Spinach simulations adhere closely to real-world experimental results~\cite{spinach_benchmarking}~\cite{spinachgoodapprox}. In the Spinach simulator, the only point where empirical values need to be provided is in the initial molecule specification step. For this specification, we use empirical data obtained through the process we have described in Fig.~\ref{DataGenerationDiagram}.

\subsection{Amino Acid Classification Dataset}
\label{amino-acid-dataset}
Using the procedure proposed in the preceding section, we develop an amino acid NMR dataset consisting of 20,000\footnote{1000 FID for each of the 20 amino acids.} simulated proton-proton TOCSY experiments. Each set of data has the dimensions of $2\times256\times256$ (2 channels of data with 256 transients each with 256 data poaints) and the whole dataset can be thought of as 46 hours of dual-channel 8k audio if we stack the transients. This is comparable in size with the VCTK~\cite{vctk} speech corpus (44 hours) but smaller than the WSJ0~\cite{wsj0} corpus (80 hours). While this is an enormous NMR dataset that would not be practical to collect physically, simulating this dataset took only around 50 hours on a system with a single Quadro P8000 GPU with 16GB RAM.

The choice to simulate amino acids is motivated by the fact that individual amino acid experiments can be linearly combined, or fed into a generative adversarial network~\cite{SimuGAN}, to approximate the spectrum of much larger proteins. Meanwhile, TOCSY experiments of proteins contain some of the most important structural information about a protein while also being challenging for a human expert to interpret, often times requiring additional NMR experiments to support its interpretation.

\section{Results}
\label{results-and-discussion}
The goal of our experimentation with using acoustic embedding models on NMR data is to test the limits of the analogy between NMR signals and audio signals while showing that it is possible to make use of acoustic signal embeddings for amino acid classification with no modifications to the acoustic embedding model. We compare the difference in performance when we use a Mel filter bank and if we use a trainable 1D convolutional encoder as shown in Table~\ref{experiment results}.

All models are trained for 10 epochs with a batch size of 20, using a cross entropy loss function and the Adam optimizer with a learning rate of 0.001. From the dataset obtained as described in~\ref{amino-acid-dataset}, we randomly select 70\% of the data to make up the training dataset, 20\% to make up the validation dataset and the remaining 10\% serves as the test dataset. This sampling is done for each amino acid, so there is no class imbalance between the different amino acids.

\begin{table}[ht]
    \setlength{\tabcolsep}{8pt} 
    \renewcommand{\arraystretch}{1.5} 
    \centering
    \caption{Classification Accuracy compared across different acoustic signal embedding models on the NMR classification dataset}
    \label{experiment results}
    \begin{tabular}{lcc}
        \hline
        \multirow{2}{*}{\textbf{Embedding Model}} & \multicolumn{2}{c}{\textbf{Encoder Model}}\\ 
         & Mel & 1D Conv\\
        \Xhline{3\arrayrulewidth}
        X-Vector     &  0.828 & 0.948 \\
        ECAPA-TDNN-S &  0.933 & 0.968 \\
        ECAPA-TDNN-L &  0.942 & \textbf{0.977} \\
        \hline
        DEEP Picker & 0.620 & 0.791 \\
        \hline
    \end{tabular}
\end{table}

To show that the 1D Convolutional Encoder is indeed able to bridge the gap between NMR and audio data, we adopt pre-trained ECAPA-TDNN-L weights from the Speechbrain Hugging Face repository and compare the results in Table~\ref{Trainable}. The pre-trained model was trained on the Vox Celeb dataset which consists of human speech.

\begin{table}[ht]
    \setlength{\tabcolsep}{8pt} 
    \renewcommand{\arraystretch}{1.5} 
    \centering
    \caption{Classification Accuracy on pre-trained ECAPA-TDNN-L}
    \label{Trainable}
    \begin{tabular}{cccc}
        \hline
        \multirow{2}{*}{\textbf{Encoder}} & \multirow{2}{*}{\textbf{Embedding Model}} & \multicolumn{2}{c}{\textbf{Weights}}\\
         & & Frozen & Trainable\\
        \Xhline{3\arrayrulewidth}
        Mel & ECAPA-TDNN-L (Pre-trained)    &  0.416 & 0.956 \\
        1D Conv & ECAPA-TDNN-L (Pre-trained) &  0.727 & 0.967 \\
        \hline
    \end{tabular}
\end{table}

\begin{figure*}[!htbp]
\begin{center}
\includegraphics[width=170mm]{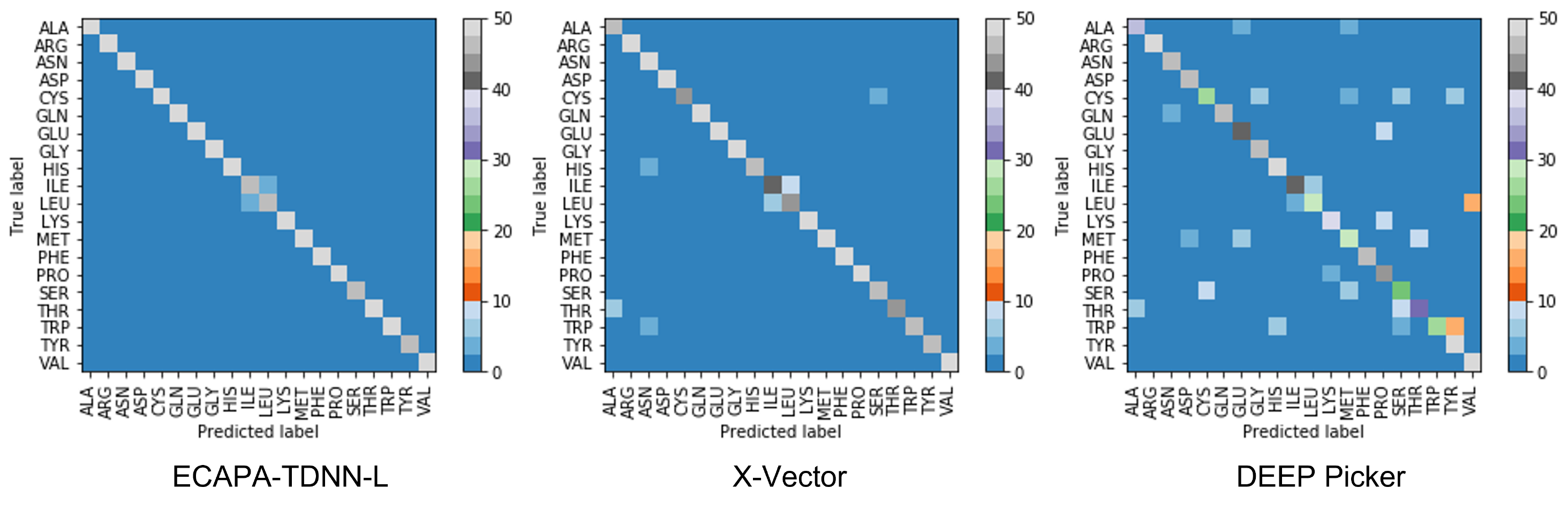}
\end{center}
\caption{The confusion matrix of the model on the test set using each of the different embedding methods. A 1D Convolutional encoder is used in all cases.}
\label{confusion-matrix}
\end{figure*}

To better understand the failure modes of our classification model and identify trends, we plot the confusion matrix of the various models making use of different embeddings in Fig.~\ref{confusion-matrix} as well as report the F1-scores of these models in Table~\ref{f1scores}. The F1-scores and confusion matrix for the model using ECAPA-TDNN-S embeddings are intentionally omitted for brevity as the results are similar to ECAPA-TDNN-L.

\begin{table}[ht]
    \setlength{\tabcolsep}{6pt} 
    \renewcommand{\arraystretch}{1.6} 
    \centering
    \caption{F1 Scores by class produced by each of the embedding models on the test set}
    \label{f1scores}
    \begin{tabular}{c|ccc}
        \hline
        \multirow{2}{*}{\textbf{Amino Acids}} & \multicolumn{3}{c}{\textbf{Encoder Model}}\\ 
         & ECAPA-TDNN-L & X-Vector & DEEP Picker\\
        \Xhline{3\arrayrulewidth}
        ALA & 0.990 & 0.911 & 0.763\\
        ARG & 0.990 & 1.000 & 0.952\\
        ASN & 0.990 & 0.926 & 0.900\\
        ASP & 1.000 & 0.990 & 0.885\\
        CYS & 0.970 & 0.896 & \textbf{0.587}\\
        GLN & 0.990 & 0.980 & 0.918\\
        GLU & 0.990 & 0.990 & 0.781\\
        GLY & 0.970 & 0.951 & 0.895\\
        HIS & 0.980 & 0.969 & 0.942\\
        \textbf{ILE} & \textbf{0.920} & \textbf{0.828} & 0.854\\
        \textbf{LEU} & \textbf{0.909} & \textbf{0.863} & 0.651\\
        LYS & 1.000 & 0.990 & 0.821\\
        MET & 0.980 & 0.990 & 0.615\\
        PHE & 1.000 & 0.990 & 0.940\\
        PRO & 1.000 & 0.980 & 0.779\\
        SER & 0.949 & 0.900 & \textbf{0.516}\\
        THR & 0.990 & 0.915 & 0.688\\
        TRP & 0.951 & 0.931 & 0.602\\
        TYR & 0.969 & 0.959 & 0.774\\
        VAL & 1.000 & 1.000 & 0.828\\
        \hline
    \end{tabular}
\end{table}

\section{Discussion}\setlength{\parskip}{1.2pt}
The experimental results reported in Table~\ref{experiment results} show that the acoustic models are able to achieve classification accuracy on the NMR dataset well above the random guess rate of 5\% for this 20-class classification problem. Importantly, the performance difference between the x-vector model and ECAPA-TDNN also shows that this classification task is non-trivial. The ECAPA-TDNN model, which typically outperforms the x-vector model on an audio dataset, also outperforms x-vector on the NMR dataset. 

Comparing the results across different encoder models, we see that performance is noticeably better when a trainable encoder is used instead of the Mel filter bank. This supports the idea that the two types of data are similar since a trainable encoder is able to map NMR data sufficiently well to an embedding space that can work well with the speaker classification model.

When the encoder model is switched to an NMR-based model, DEEP Picker~\cite{deeppick}, we see that the performance of the model is reduced. The best performing ECAPA-TDNN-L embedding model achieves a 23\% relative improvement in performance compared to when DEEP Picker is used. This highlights the effectiveness of using a ready-made audio classification model for NMR tasks. We note however that this is not necessarily a fair comparison as the DEEP Picker model has a smaller model size of 0.4M parameters while the ECAPA-TDNN-L model has 104M parameters (both models using the 1D Convolutional Encoder). While it might be possible to increase the number of parameters of DEEP Picker by increasing the kernel sizes and number of filters, there is a limit to scaling up the number of parameters without fundamentally changing model architecture. In the absence of larger NMR classification models, DEEP Picker presents the best available comparison.

Another way to view the comparison between DEEP Picker and ECAPA-TDNN-L is as a validation of our approach of using acoustic signal embedding models for NMR classification. Given that larger models often lead to better performance if the model architecture is sufficiently well designed, the methods we would use to improve the performance of DEEP Picker would be very similar to acoustic models. For example, we would add residual connections as we increase the depth of the model or include squeeze and excitation blocks to further strengthen the representational power of the model. The fact that ECAPA-TDNN-L performs better than DEEP Picker shows that the effectiveness of these innovations in acoustic signal processing carries over to NMR data.

The experimental results reported in Table~\ref{Trainable} highlight the importance of the 1D convolutional encoder. The model performs poorly when using a Mel filter bank encoder, even if a 42\% accuracy is still well above random guessing. An interesting result considering that the frozen pre-trained embedding model has not seen NMR data before. Nevertheless, when we swap the Mel filter bank for a 1D convolutional encoder, we see a near doubling of accuracy to 73\% even in the experiment where the embedding weights are frozen. This further supports the idea that NMR and acoustic data are similar since the encoder needs to map NMR data into an acoustic feature space in order to work well with a frozen acoustic signal embedding model. When the pre-trained weights are made trainable and fine-tuned on the NMR dataset, we see performance in line with Table~\ref{experiment results} as expected. 

Looking at the confusion matrix of the different models in Fig.~\ref{confusion-matrix}, we can see that the models consistently make classification errors on the amino acids ILE and LEU, even for the audio embedding models which have better overall accuracy. For the DEEP Picker model errors happen more broadly across different amino acids due to its overall poorer performance, although ILE and LEU still have a substantial contribution. Looking at the detailed F1 scores of each of the amino acids across each of the models in Table~\ref{f1scores}, we can see more clearly that these two amino acids make up the two lowest F1 scores for both the ECAPA-TDNN-L and x-vector models. 

\begin{figure}[!htbp]
\begin{center}
\vspace*{-3pt}
\includegraphics[width=80mm]{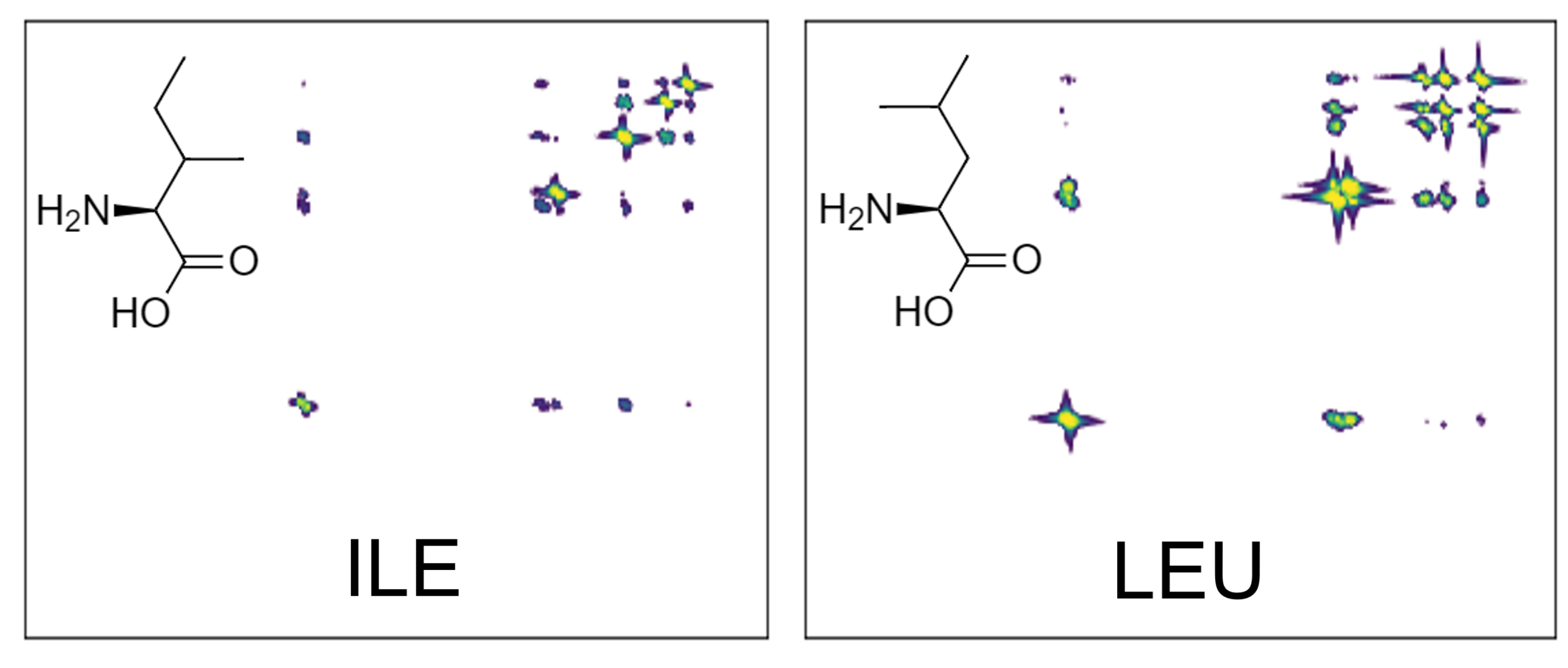}
\end{center}
\caption{Comparison of the 2D interferogram of the ILE and LEU signals. The molecular structures of these two amino acids are very similar and their NMR signals are accordingly very difficult to differentiate.}
\label{ile-leu}
\end{figure}

To understand why the model has a more difficult time classifying the ILE and LEU, we plot the 2D NMR interferogram of a sample of each of these amino acids from our dataset in Fig.~\ref{ile-leu}. We can see clearly in the frequency domain that these two amino acids have very similar signatures. This is due to the fact that they have very similar chemical structures, which are also drawn at the side of the spectra in Fig.~\ref{ile-leu}. Both amino acids have side chains with an identical number of hydrogen atoms, which means that they will have an identical number of resonances. The difference in structure between the amino acids is subtle, differing only in the position of one of their $CH_{3}$ groups, so the resonances will also be found in very similar positions on the interferogram, making it very difficult to distinguish between these two amino acids. Despite this, the ECAPA-TDNN-L model is able to achieve F1-scores of 0.920 and 0.909 on ILE and LEU respectively, highlighting the effectiveness of the acoustic signal embeddings.

\section{Conclusion}
The development of NMR has always been in step with improvements in computing power and better algorithms. Many of the analyses commonly used everyday in the NMR lab rely on a host of algorithms, from locking magnetic fields in the instrument to software used for protein peak assignments, which have been possible only because cheap and abundant computing. Considering this history, deep learning appears to be a natural next step for the development of NMR techniques. 

In pursuit of this, we have developed a methodology for creating a large, realistic simulated NMR dataset. We have then shown that with a larger simulated dataset, we are able to plug-and-play models originally designed for audio signal processing and achieve good performance. As a proof-of-concept, we have simulated a large amino acid NMR classification dataset and made use of audio speaker embeddings from speaker verification systems to perform classification. By using an unmodified acoustic signal embedding model we were able to achieve a 23\% relative improvement in performance compared to an existing NMR-based model, which demonstrates that model architecture strategies that are used in acoustic models can also work on NMR data.

\end{document}